\pdfoutput=1 
\documentclass[a4paper,11pt]{article}
\title{An intrinsic mobility ceiling of Si bulk}
\author{Nuria Garcia-Castello\footnote{nuriagarcia@ub.edu}, Joan Daniel Prades, and Albert Cirera \\
\small MIND-IN$^2$UB, Departament of Electronics, Universitat de Barcelona,\\\small C/Mart\'i i Franqu\`es 1, E-08028 Barcelona, Spain }
\usepackage{graphicx}
\usepackage{bm}
\date{}

\begin{document}

\maketitle

\begin{abstract}
We compute by Density Functional Theory-Non Equilibrium Green Functions Formalism (DFT-NEGFF) the conductance of bulk Si along different crystallographic directions. We find a ceiling value for the intrinsic mobility of bulk silicon of $8.4\cdot10^6 cm^2/V·s$. We suggest that this result is related to the lowest effective mass of the $<$001$>$ direction.
\end{abstract}

\begin{figure}[htbp]
	\centering
		\includegraphics[width=0.3\textwidth]{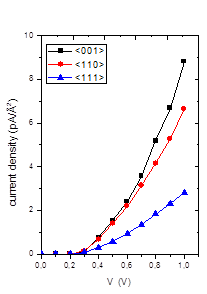}
	\caption{Comparison of the I(V) characteristic (I in this case is the current per unit of transversal area) of the three directions of cubic diamond $<$001$>$, $<$110$>$ and $<$111$>$.}
	\label{fig:1_intensity}
\end{figure}

Silicon is the most widely used material in current electronic devices. Due to downscaling, much work has been done on nanostructures such as nanowires. Their structural and electrical properties have been studied experimentally and theoretically \cite{Schmidt} and devices have been made with them \cite{LogicGates}. Since these new structures are to be used in electronic devices, an understanding of their electronic transport behaviour is essential. \\
Carrier mobility is an important transport property of a semiconductor. It reflects the capacity of the charge to move in the material and to respond to field changes \cite{transport2}. \\
Due to the challenges of current \textsl{ab initio} computations, transport properties have been studied by this method only in very low dimensional systems \cite{SiNWteo,SiNWpetit,NEGFF}. The most common way to examine transport and mobility in bulk silicon has been to solve the Boltzmann equation using MonteCarlo methodology with electron-phonon interaction as a scattering mechanism \cite{Pop,Akkerman,driftvelocity,Ungersboeck,Schenk}. To our knowledge, no theoretical approach based on fully atomistic \textsl{ab initio} methods has been used to study bulk behavior. \\
Materials in devices are normally used in an anisotropic way: they are grown along a certain axis and one direction of transport becomes predominant, for example, silicon nanowires (SiNWs). The transport properties are expected to be sensitive to the crystal orientation, due to the anisotropy of band structure of bulk silicon. Thus, a detailed understanding of the electrical properties as a result of the different growth conditions is required in order to assess the impact of the crystal structure on transport properties and to determine which orientation is best suited to the engineering of high performance thick SiNW-based devices.\\
In this Letter, we present a theoretical study of the electron transport properties of bulk silicon as an approximation to the transport in thick nanodevices, where the bulk properties are still predominant over confinement effects \cite{reviewSiNWs}. For the sake of completeness, the most significant growth directions of SiNWs, namely cubic diamond phase oriented along $<$001$>$, $<$110$>$ and $<$111$>$ axis, have been considered. We have evaluated the intrinsic electron mobility due to elastic scattering of the electrons inside the crystal. We compare the values with current experimental results from the literature and with the corresponding results for small SiNWs, finding a ceiling of the value of bulk silicon mobility depending on the crystallographic direction which is lower than the reported value for SiNWs \cite{Yu,SiNWexp,Ng}. The authors suggest that surface states may be responsible for the measured conduction. \\
Here the electronic transport is studied inside an open-boundary bulk system described by first-principles field within Non Equilibrium Green Functions Formalism (NEGFF) \cite{Datta,Haug}. We performed Density Functional Theory (DFT) calculations with a numerical atomic orbital basis set \cite{siesta}. We used a double-$\zeta$ polarized basis set, with the generalized gradient approximation \cite{PBE} for the exchange-correlation functional in the PBEsol parametrization \cite{PBEsol}. Real space mesh cut-off and k-grid values were set to converge total energy values within 0.1 meV and maximum absolute forces over atoms are smaller than 0.004 eV/Ang in all cases. \\
The electronic transport properties were studied with Non Equilibrium Green Functions Formalism, within the Keldysh formalism \cite{Keldish}, based on DFT as implemented in TranSIESTA code \cite{TranSIESTA}. It computes the electronic transport properties due to elastic scatter in a ballistic regime of a nanoscale system confined in a central scattering region (C) in contact with two semi-infinite left (L) and right (R) electrode regions at different electrochemical potentials. The bias voltage $V_b$ between them corresponds to the electrochemical potential difference between the left ($\mu_L$) and the right ($\mu_R$) electrodes, $eV_b=\mu_L-\mu_R$. In our case, both electrodes and the contact region were made of Si in the same crystallographic orientation. \\
The density matrix from the incoming scattering states of the left (l) and right (r) electrode is
	\[
D(\vec{x},\vec{y})=\sum_l{\psi_l(\vec{x})\psi_l^*(\vec{y})n_F(\epsilon_l-\mu_L)}+ \sum_r{\psi_r(\vec{x})\psi_r^*(\vec{y})n_F(\epsilon_r-\mu_R)}, 
\]
where the scattering states starting in the left electrode are
\[
	\psi_l(\vec{x})=\psi_l^0(\vec{x})+\int{d\vec{y}G(\vec{x}\vec{y})V_L(\vec{y})\psi_l^0(\vec{y})}. 
\]
These are generated from the unperturbed incoming states of the uncoupled, semi-infinite electrode $\psi_l^0$, using the retarded Green's function G of the coupled system \cite{TranSIESTA} and the interaction between the left electrode and the contact region $V_L$.\\
The current I through the contact region was calculated using the Landauer-Buttiker formula \cite{transport},
\[
I(V)=G_0\int_{-\infty}^{\infty}{d\epsilon[n_F(\epsilon-\mu_L)-n_F(\epsilon-\mu_R)]T(\epsilon, V_b)},
\]
where $G_0 = 2e^2/h$ is the unit of quantum conductance, $n_F$ is the electrode distribution function and $T(\epsilon,V_b)$ is the transmission probability of electrons incident at an energy $\epsilon$ through the device under the potential bias $V_b$.\\
We have computed the I(V) characteristic (Fig. \ref{fig:1_intensity}) in each direction and evaluated the intrinsic conductance.\\

\begin{figure}[htbp]
 \centering
	\includegraphics[width=0.3\textwidth]{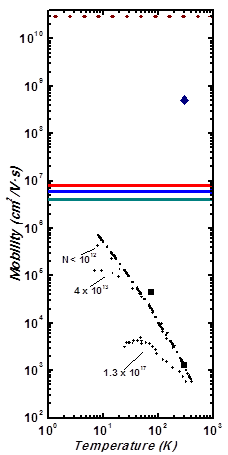}
		\caption{Extract from ref. \cite{transport2}. Experimental mobility of electrons in Si as a function of temperature with different impurity concentrations (point symbols). The dash-dot line indicates the pure lattice mobility obtained with Monte-Carlo simulation \cite{driftvelocity}. Our computed mobilities (continuous lines: red for $<$001$>$, blue for $<$110$>$ and green for $<$111$>$) are compared with experimental values for Si bulk (point symbols \cite{transport2} and square symbol \cite{shalimova}) and experimental \cite{SiNWexp} (diamond symbol) and theoretical (with NEGFF) \cite{Ng} (dot-dot line) values of thin SiNWs.}
	\label{fig:2_mobility}
\end{figure}

We propose to calculate the corresponding electron mobility $\mu$ of the electrons along the different directions of bulk Si. Assuming an intrinsic concentration of carriers n corresponding to a temperature of 300K, the corresponding mobility $\mu=\sigma/e·n$ (where $\sigma$ is the associated conductivity computed as the conductance across the corresponding unit cell in each direction) is shown in Fig. \ref{fig:2_mobility}. \\

The mobility in bulk semiconductors as we go down from room temperature increases at first due to the suppression of phonon scattering \cite{Datta}. But it does not increase any further once the phonon scattering is small enough so that impurity scattering becomes the dominant mechanism. With undoped samples we can obtain higher mobilities, because we have suppressed inelastic scattering centers (Fig. \ref{fig:2_mobility}).  \\
In agreement with these considerations our computed mobility is greater than experimental results and thus constitutes a ceiling value for them. TranSIESTA code only takes into account the transport due to elastic scatter inside the crystal, without considering the contribution of inelastic scatter due to phonons and impurities. Despite current advances \cite{inelastica}, up to now there is no reliable way to compute inelastic scattering in this framework. On the other hand, our calculated conductivities are lower than experimental ones in SiNWs because in bulk the surface states are not taken into account, and they seem to play an important role in transport properties in very thin SiNW \cite{Schmidt,SiNWexp,Ng}. Thus, our result becomes a threshold to distinguish bulk conductivity from conductivity affected by nanoscale effects.\\
Electrons in silicon occupy several valleys which have different orientations. The effective mass associated with these valleys gives the response of the electrons to an external field; the electrons with lower effective mass are faster, and the direction of these valleys is the preferred direction of the electron transport. The results are similar to the experiments \cite{costato} to increase the mobility due to an application of a uniaxial stress along a direction such that the valleys presenting the lower effective mass in the direction of the field have less energy than the other valleys. Electrons in equilibrium will have greater tendency to occupy these "faster valleys", and the overall mobility will increase.\\
We have calculated the effective mass (see Table \ref{table1}), with respect to the free electron mass $m_0$, around the conduction band minimum (CBM) in each direction using the expression $\Delta E_c(k)= \hbar^2k^2/2m^*$, where m* is the electron effective mass around the CBM. We fit a quadratic of the form $y=mx^2$, then we solve $m=\hbar^2/2m^*$. For bulk Si in $<$001$>$ direction, we calculate m* to be 0.884 $m_0$. This value corresponds to the longitudinal effective mass of electrons in bulk silicon and it is close to the reported experimental values \cite{mass_effective}. In order to check the calculations we computed the electron transversal effective mass and the light hole effective mass, and the values were 0.194 $m_0$ and 0.163 $m_0$, respectively. They are close to the reported experimental values \cite{mass_effective2}.\\
\begin{table}[htbp]
	\centering
		\begin{tabular}{l c c c}
		\hline
		\	 & $<$001$>$ & $<$110$>$ & $<$111$>$ \\
			 \hline
		electron effective mass & 0.884 $m_0$ & 1.198 $m_0$ & 1.575 $m_0$ \\
		mobility $(cm^2/V·s)$ & $8.4\cdot10^6$ & $6.04\cdot10^6$ & $3.99\cdot10^6$ \\	 
		\hline
		\end{tabular}
	\caption{Electron effective mass and mobility for bulk Si oriented in $<$001$>$, $<$110$>$ and $<$111$>$ direction. }
	\label{table1}
\end{table} 
\\
In conclusion, we have performed a theoretical study of the mobility of bulk Si based on the description with NEGFF of the electronic transport. We have evaluated different crystallographic directions and found the $<$001$>$ the best for electrons in bulk Si due to the absolute minimum of the lowest conduction bands in the band diagram and the lower corresponding effective mass. We have performed the calculations inside the framework of Landauer-Buttiker formalism and found an intrinsic conductance due to elastic scattering of electrons inside silicon bulk, and we can associate a mobility value to it. In this sense, our computed values are a ceiling (i.e. the maximum possible mobility that the material could achieve). With the contribution of inelastic scatter due to phonons and impurities, the mobility is reduced. Moreover, our result becomes a threshold between bulk conductivity and conductivity affected by nanoscale effects. While the Non Equilibrium Green Functions Formalism is currently used to study transport in very small nanostructures, we have used it in huge configurations (bulk), showing that it is a very powerful technique for evaluating properties of many-body systems. \\
\\
N. G.-C. acknowledges the Spanish MICINN for her PhD grant in the FPU program. A.C. acknowledges the support from ICREA academia program. The authors thankfully acknowledge the computer resources, technical expertise and assistance provided by the Barcelona Supercomputing Center - Centro Nacional de Supercomputaci\'{o}n.


\begin{thebibliography}{99}
\bibitem{Schmidt} V. Schmidt, J.V. Wittemann, and U. G\"{o}sele,  Chem. Rev. \textbf{110}, 361 (2010).
\bibitem{LogicGates} Y. Huang, X. Duan, Y. Cui, L.J. Lauhon, K. Kim, and C.M. Lieber, Science \textbf{294}, 1313 (2001).
\bibitem{transport2} C. Jacoboni, \textsl{Theory of Electron Transport in Semiconductors} (Springer,Berlin,2010).
\bibitem{SiNWteo} M.P. Persson, A. Lherbier, Y.-M. Niquet, F. Triozon, and S. Roche, Nano Lett. \textbf{8}, 4146 (2008).
\bibitem{SiNWpetit} R. Rurali and N. Lorente, Phys. Rev. Lett. \textbf{94}, 026805 (2005).
\bibitem{NEGFF} N. Seoane, A. Martinez, A.R. Brown, and A. Asenov, Proceedings of the 2009 Spanish Conference on Electron Devices (2009).
\bibitem{Pop} E. Pop, R.W. Dutton, and K.E. Goodson, J. Appl. Phys. \textbf{96}, 4998 (2004).
\bibitem{Akkerman}	A. Akkerman, M. Murat, and J. Barak, J. Appl. Phys. \textbf{106}, 113703 (2009).
\bibitem{driftvelocity}	C. Canali, C. Jacoboni, F. Nava, G. Ottaviani, and A. Alberigi-Quaranta, Phys. Rev. B \textbf{12}, 2265 (1975).
\bibitem{Ungersboeck}	E. Ungersboeck, S. Dhar, G. Karlowatz, V. Sverdlov, H. Kosina, and S. Selberherr, IEEE Trans. Electron Devices \textbf{54}, 2183 (2007).
\bibitem{Schenk}	A. Schenk, J. Appl. Phys. \textbf{79}, 814 (1996).
\bibitem{reviewSiNWs} P.R. Bandaru and P. Pichanusakorn, Semicond. Sci. Technol. \textbf{25}, 024003 (2010).
\bibitem{Yu} J.-Y. Yu, S.-W. Chung, and J.R. Heath, J. Phys. Chem. B \textbf{104}, 11864 (2000).
\bibitem{SiNWexp} J. Bauer, F. Fleischer, O. Breitenstein, L. Schubert, P. Werner, U. Gösele, and M. Zacharias, Appl. Phys. Lett. \textbf{90}, 012105 (2007). 
\bibitem{Ng} M.-F. Ng, L. Shen, L. Zhou, S.-W. Yang, and V. Tan, Nano Lett. \textbf{8}, 3662 (2008).
\bibitem{Datta} S. Datta, \textsl{Electronic Transport in Mesoscopic Systems} (The Press Syndicate of the University of Cambridge, Cambridge, 1995).
\bibitem{Haug} H. Haug and A.-P. Jauho, \textsl{Quantum Kinetics in Transport and Optics of Semiconductors} (Springer-Verlag, Berlin, 1996).
\bibitem{siesta} J. M. Soler, E. Artacho, J.D. Gale, A. Garc{\'\i}a, J. Junquera, P. Ordej\'{o}n, and D. S\'{a}nchez-Portal, J. Phys. Condens. Matter \textbf{14}, 2745 (2002).
\bibitem{PBE} J. P. Perdew, K. Burke, and M. Ernzerhof, Phys. Rev. Lett. \textbf{77}, 3865 (1996).
\bibitem{PBEsol} J. P. Perdew \textsl{et al.}, Phys. Rev. Lett. \textbf{100}, 136406 (2008).
\bibitem{Keldish} R. van Leeuwen \textsl{et al.}, Lect. Notes Phys. \textbf{706}, 33 (2006).
\bibitem{TranSIESTA} M. Brandbyge, J.-L. Mozos, P. Ordej\'{o}n, J. Taylor, and K. Stokbro, Phys. Rev. B \textbf{65}, 165401 (2002)
\bibitem{transport} D.K. Ferry, S.M. Goodnick, and J. Bird, \textsl{Transport in Nanostructures} (Cambridge University Press, New York, 2009)
\bibitem{shalimova} K. V. Shalimova, \textsl{Phyisics of Semiconductors} (Editorial Mir, Moscou, 1982).
\bibitem{inelastica} T. Frederiksen, M. Brandbyge, N. Lorente, and A.-P. Jauho, Phys. Rev. Lett. \textbf{93}, 256601 (2004).
\bibitem{costato} M. Costato and L. Reggiani, Lettere al Nuovo Cimento \textbf{4}, 848 (1970).
\bibitem{mass_effective} G. Harbeke, O. Madelung, and U. Rössler, \textsl{Numerical Data and Functional Relationships in Science and Technology} (Springer-Verlag, Landoldt-Börnstein, Berlin, 1982).
\bibitem{mass_effective2} R. N. Dexter, Phys. Rev. \textbf{96}, 223 (1954).
\end{thebibliography}

\end{document}